\documentclass[letterpaper, 10 pt, conference]{ieeeconf}

\usepackage{blindtext, graphicx}
\usepackage{color}
\usepackage{amsmath,amssymb}
\usepackage{subcaption}
\usepackage{siunitx}
\usepackage{dblfloatfix}

\IEEEoverridecommandlockouts      
\title{\LARGE \bf Stability and Control of Ad Hoc DC Microgrids}
\author{Julia A. Belk, Wardah Inam, David J. Perreault, Konstantin Turitsyn
\thanks{J.A. Belk, W. Inam, and D.J. Perreault are with the Electrical Engineering and Computer Science Department and K. Turitsyn is with the Mechanical Engineering Department, Massachusetts Institute of Technology, Cambridge, MA, 02139 USA, e-mails: \{jabelk, wardah, djperrea, turitsyn\}@mit.edu.
}
}

\begin{document}

\maketitle
\thispagestyle{empty}
\pagestyle{empty}

\begin{abstract}
Ad hoc electrical networks are formed by connecting power sources and loads without pre-determining the network topology. These systems are well-suited to addressing the lack of electricity in rural areas because they can be assembled and modified by non-expert users without central oversight. There are two core aspects to ad hoc system design: 1) designing source and load units such that the microgrid formed from the arbitrary interconnection of many units is always stable and 2) developing control strategies to autonomously manage the microgrid (i.e., perform power dispatch and voltage regulation) in a decentralized manner and under large uncertainty. To address these challenges we apply a number of nonlinear control techniques---including Brayton-Moser potential theory and primal-dual dynamics---to obtain conditions under which an ad hoc dc microgrid will have a suitable and asymptotically stable equilibrium point. Further, we propose a new decentralized control scheme that coordinates many sources to achieve a specified power dispatch from each. A simulated comparison to previous research is included.
\end{abstract}

\section{Introduction}

More than one billion people do not have electricity access \cite{iea15}, largely because of insufficient centralized power systems in developing countries. The need for electricity in remote and rural areas and the evolving demands on the existing bulk power infrastructure have driven extensive development of microgrids in recent years. Microgrids naturally incorporate distributed renewable sources and are inherently decentralized. However, designing and installing a microgrid to electrify an off-grid community typically requires specialized planning. Electrical networks which could be formed by the ad hoc interconnection of modular power sources and loads by non-specialist users would remove barriers to energy access, allowing decentralized electricity markets to proliferate in an unprecedented manner. 

Broadly, the critical challenges of ad hoc microgrids are:

\begin{enumerate}
    \item the microgrid components (power sources, loads, and lines) should be designed so that \textit{any} network formed by connecting many units \textit{always} has an appropriate and stable equilibrium point,
    \item the sources should be controlled in a coordinated and decentralized manner to manage power dispatch, and
    \item the system should function entirely autonomously under significant uncertainty regarding the network configuration and power supply and demand.
\end{enumerate}

To evaluate stability, traditional power system operators use computationally intensive simulations and empirical testing of the pre-determined network topology. By contrast, ad hoc networks present unique challenges because the network topology is not specified. Instead, analytic constraints on the individual source, load, and line units---\textit{independent from the interconnection structure}---must be developed such that, when satisfied, they guarantee the existence, feasibility, and stability of an equilibrium point. This is different from previous stability analyses in power electronics, which have relied on impedance measurements, used overly simplified network models, or placed unrealistic constraints on the network configuration \cite{CriteriaReview,ROM}. An additional complication is the widespread and increasing use of power electronic devices, which draw constant power at their inputs to regulate their outputs. The negative incremental impedance ($\partial v/\partial i$) of these loads has a well-documented destabilizing effect on power systems \cite{cplinstability}. The effect of constant power loads on stability has attracted recent interest in the controls community (\cite{Sanchez:2013gl,Bolognani:2015ek,SimpsonPorco:2015hp,Cezar:2015io,Barabanov:2016ki}) but has not yet been analyzed in the context of ad hoc systems. 

To plan and coordinate power dispatch, traditional ac power system operators compute the optimal dispatch based on the marginal costs of the generators, and each generator realizes its assigned power output by controlling the phase angle between its output voltage and current. Ad hoc systems again present unique challenges because of the combination of the need for decentralized computation and control and the uncertainty in power supply and demand. In dc systems powered by voltage-source converters, each source realizes its assigned power output by controlling its voltage level. A set of control techniques, organized into a hierarchy, has been widely used in ac and dc microgrids to accomplish these objectives \cite{GuerreroHierarchy}, and economically optimal dispatch strategies for dc microgrids have been considered \cite{Zhao:2015eu}, but again, not in the context of nonlinear ad hoc systems.

In this paper, we develop design-friendly conditions on both the total system load and the individual source, load, and line units under which microgrids composed of these units will have a suitable and stable equilibrium point. Our conditions are summarized by Eqs. \eqref{eq:result2}, \eqref{eq:result3}, and \eqref{eq:result1}. They are more flexible and better suited to a priori device design than previous stability criteria, and rely on techniques from nonlinear control theory \cite{Brayton:1964gr,Jeltsema:2003jz,Jeltsema:2009jd,Feijer:2010ia}. We also propose a decentralized control strategy which achieves precise power sharing and voltage regulation, and include a simulated comparison to a well-known power electronics benchmark. Throughout, we discuss how previously developed mathematical structures can be related to the analysis of stability and control of ad hoc systems and highlight opportunities and challenges related to this emerging class of power systems.

\begin{figure*}[b!]
\begin{subfigure}{.271\textwidth}
  \centering
  \includegraphics[width=\linewidth]{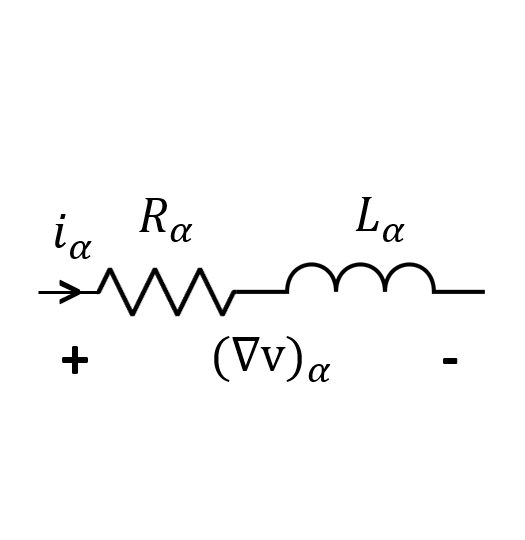}
  \caption{Line with impedance $R_{\alpha} + j\omega L_{\alpha}$. $L_{\alpha}/R_{\alpha} = \tau_{\alpha}$.\newline}
  \label{fig:line}
\end{subfigure}
\hfill
\begin{subfigure}{.314\textwidth}
  \centering
  \includegraphics[width=\linewidth]{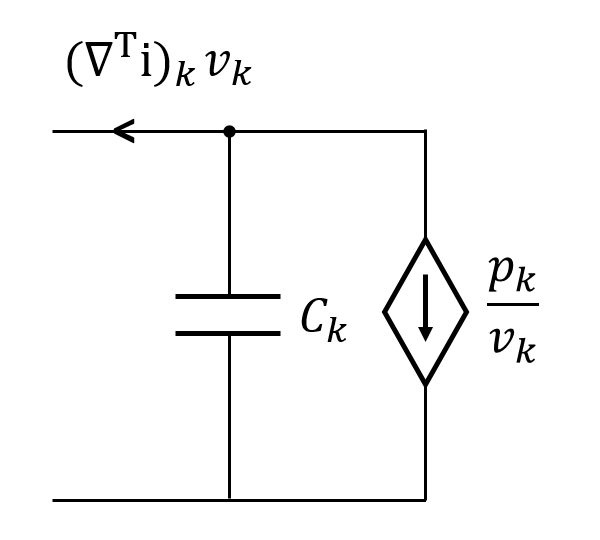}
  \caption{Ideal model of load and associated power electronics: constant power load with parallel stabilizing capacitance $C_k$.}
  \label{fig:load}
\end{subfigure}
\hfill
\begin{subfigure}{.314\textwidth}
  \centering
  \includegraphics[width=\linewidth]{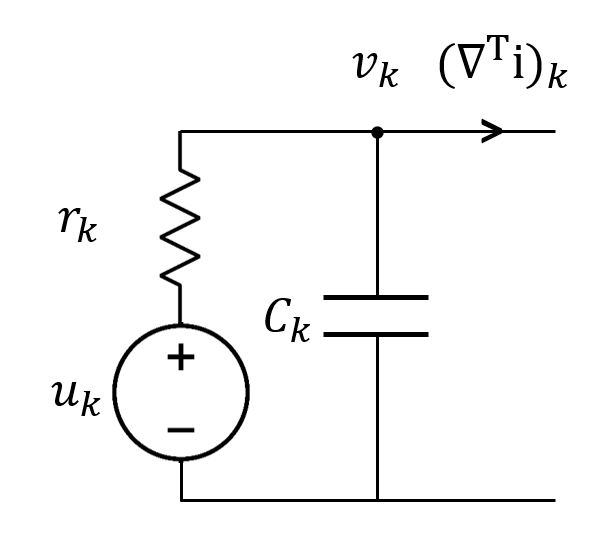}
  \caption{A voltage source converter with droop resistance $r_k$ and parasitic parallel capacitance $C_k \to 0$.}
  \label{fig:source}
\end{subfigure}
\caption{Representations of lines, loads, and sources in our dc microgrid.}
\label{fig:models}
\end{figure*}

\section{Models and Notation}\label{sec:models}

In this section we present models for the interconnecting lines, power electronic loads, and voltage source converters which are analytically tractable and can be adapted to describe many networks. These are summarized in Fig. \ref{fig:models} and are based on a previously presented ad hoc microgrid \cite{ugrid}.

\subsection{Network Structure}

The electrical network is described as a weighted, directed graph $({\cal V,\cal E})$ with a total of  $|{\cal V}| = n$ nodes (buses) and $|{\cal E}| = m$ edges (lines). A power source or load is attached to each node and the models for each element are depicted in Fig. \ref{fig:models}. The electric state of the system is described by the voltage and current vectors $v \in \mathbb{R}^n$ and $i \in \mathbb{R}^m$. The topology of the graph is defined by an incidence matrix $\nabla \in \mathbb{R}^{m\times n}$, such that applying $\nabla$ to the voltage vector results in a potential drop across line $\alpha$ with source bus $s_\alpha$ and target bus $t_{\alpha}$:
\begin{equation}\label{Eq:vdrop}
    (\nabla v)_\alpha = v_{s_\alpha} - v_{t_\alpha}\,.
\end{equation}

\noindent Similarly, applying $\nabla^\top$ to the current vector provides the total current flowing out of each node:
\begin{equation}\label{Eq:isum}
 (\nabla^\top i)_k = \sum_{\{\alpha:\,\, s_\alpha = k\}} i_\alpha - \sum_{\{\alpha:\,\, t_\alpha = k\}} i_\alpha\,.
\end{equation}


\subsection{Power Lines (Edges)}

Each power line is associated with a graph edge $\alpha \in {\cal E}$ and is characterized by an inductance $L_\alpha$ and a resistance $R_\alpha$, as shown in Fig. \ref{fig:line}. Each line has time constant $\tau_\alpha = L_\alpha/R_\alpha$, and each $i_{\alpha}$ is described by:

\begin{equation}
    L_\alpha \dot i_\alpha = - R_\alpha i_\alpha + (\nabla v)_\alpha, \quad \alpha \in {\cal E}\,.
\end{equation}

\subsection{Sources/Loads (Nodes)}

Each bus has either a load or a source attached to it. We denote the subset of vertex indices corresponding to loads as ${\cal V}_l \subset {\cal V}$ with $|{\cal V}_l| = n_l$ and the subset of source indices as ${\cal V}_s \subset {\cal V}$ with $|{\cal V}_s| = n_s$. Load $k$ is represented by the parallel connection of a capacitance $C_k$ and a constant power load drawing power $p_k$. In general, constant power loads represent perfectly-regulated power converters, and hence are conservative and general models which can be used to describe many power electronic devices. The capacitor across the input of the power converter is a standard feature of these converters, and is critical for system stability \cite{Cezar:2015io}. Each load voltage is described by:

\begin{equation}
    C_k \dot{v}_k = -\frac{p_k}{v_k} - (\nabla^\top i)_k, \quad k \in {\cal V}_l\,.
\end{equation}

Sources are represented as a voltage source with value $u_k$ in series with a resistance $r_k$. Both $u_k$ and $r_k$ are internal control parameters and can be varied independently---an overview is provided in Section \ref{sec:control} and in \cite{GuerreroHierarchy,GuerreroDroop}. To simplify our notation, a parallel capacitor is also included in the source model. Unlike the load capacitor, this is not typically present in power converters, so in our final results we will take the limit $C_k \to 0$ for the source buses. Each source voltage is described by:

\begin{equation}
    C_k \dot{v_k} = \frac{u_k - v_k}{r_k} - (\nabla^\top i)_k\,, \quad k \in {\cal V}_s\,,
\end{equation}
which reduces to 
\begin{equation}\label{eq:alg}
    v_k = u_k - r_k (\nabla^\top i)_k    
\end{equation}
as $C_k \to 0$.

Finally, we introduce a simplified control method for $u_k$, mainly to demonstrate how the stability of source voltage control strategies can be rigorously analyzed:

\begin{equation} \label{eq:dudt}
    C_u \dot{u}_k = \frac{v_{\mathrm{ref}} - v_k}{r_k}\,, \quad k \in {\cal V}_s\,,
\end{equation}
where $v_{\mathrm{ref}}$ is the nominal network voltage and $C_u$ is a control parameter (chosen instead of the more conventional $\tau_u$ for dimensional convenience). This control strategy is not typically used---a coordinated strategy for practical use is presented in Section \ref{sec:control}.

\section{Dynamics and Stability}\label{sec:dynamics}

In this section we analyze the stability of arbitrary interconnections of sources, loads, and lines, under the assumption that each source is controlled according to Eq. \eqref{eq:dudt}. This extends our previous, linear analysis \cite{compel} to include nonlinear constant power load models and to apply to more general networks---specifically, those with multiple $\tau_\alpha$ values.

We formulate the dynamical equations in a structured way so that we can apply the classic results by Brayton and Moser regarding the stability of nonlinear circuits \cite{Brayton:1964gr} to ad hoc networks. This representation allows us to find a Lyapunov function which certifies the asymptotic stability of an equilibrium point. In the next section, we reduce the derived linear matrix inequalities to design-friendly constraints on the total load and individual sources, loads, and lines.

\subsection{Brayton-Moser Representation of the System}

We first rewrite the equations from Section \ref{sec:models} in vector form. We introduce the diagonal resistance matrix $R = \mathrm{diag}[R_\alpha] \in \mathbb{R}^{m\times m}$ for the line resistances, the diagonal capacitance matrix $C = \mathrm{diag}[C_k] \in \mathbb{R}^{n\times n}$ for the source and load capacitances, the resistive content of the lines ${\cal R}_0(i)$, and the resistive co-content of the sources and loads ${\cal G}_0$ \cite{Jeltsema:2009jd}:

\begin{equation}
	{\cal R}_0(i) = \frac{1}{2}i^\top R i\,,
\end{equation}
\begin{equation}
	{\cal G}_0(v,u) =   \sum_{k\in {\cal V}_l} p_k \ln{v_k} + \sum_{k\in{\cal V}_s} \frac{v_k^2 /2 + (v_{\mathrm{ref}}-v_k) u_k}{r_k}\,.
\end{equation}

These definitions allow us to write the voltage and current equations as:
\begin{subequations}\label{eq:main}
\begin{align}
     L \dot{i}  = &
     - \partial_i {\cal R}_0(i) + \nabla v\,, \label{didt} \\
      C \dot{v} = & -\partial_v {\cal G}_0(v,u) - \nabla^\top i\,, \\
      C_u \dot{u} = &  \partial_u {\cal G}_0(v,u)\,.
\end{align}
\end{subequations}

We now introduce the Brayton-Moser (BM) Potential ${\cal P}_0$ and the matrix ${\cal Q}_0$ \cite{Brayton:1964gr,Jeltsema:2003jz}:

\begin{equation}
	{\cal P}_0 =  {\cal G}_0(v) - {\cal R}_0(i) + i^\top\nabla v\,,
\end{equation}
\begin{equation}
	{\cal Q}_0 =
    \begin{bmatrix}
	-L & 0\\
    0 & C 
\end{bmatrix}\,.
\end{equation}

These allow us to represent the current and voltage equations together with a single state vector $x = [i^\top, v^\top]^\top$ that evolves according to quasi-gradient dynamics of the form:
\begin{equation}\label{eq:quasigrad}
 {\cal Q}_0 \dot{x} = -\partial_x {\cal P}_0\,.
\end{equation}

Although this formulation reveals the structure of the underlying dynamics, it is not sufficient to certify the stability of the system, because the natural Lyapunov function candidate ${\cal P}_0(x)$ is neither convex nor sign definite. However, we can use these quantities to define a closely related BM potential ${\cal P}$ and resistive co-content ${\cal G}$:
\begin{align}\label{eq:newp}
	{\cal P}(x, u) 
	&= \frac{\tau_{max}}{2}\begin{bmatrix}
     \partial_i {\cal P}_0 \\
     \partial_v {\cal P}_0 
    \end{bmatrix}^\top
    \begin{bmatrix}
    L & 0 \\
    0 & C   
    \end{bmatrix}^{-1}
    \begin{bmatrix}
     \partial_i {\cal P}_0 \\
     \partial_v {\cal P}_0 
    \end{bmatrix} + {\cal P}_0\,,  \nonumber \\
    & = \frac{1}{2}\dot{i}^\top \left[\tau_{max} L - L R^{-1} L\right] \dot{i} \nonumber \\
    & \qquad + \frac{\tau_{max}}{2}\dot{v}^\top C \dot{v} + {\cal G}(v, u)\,. \\
    \mathcal{G}(v, u)
    &= {\cal R}_0\left(R^{-1}\nabla v\right) + {\cal G}_0(v, u)\,,  \nonumber \\
    &= \frac{1}{2} v^\top\nabla^\top R^{-1}\nabla v  + {\cal G}_0(v, u)\,.
\end{align}
Here $\tau_{max}$ denotes the maximal time-constant of all the lines: $\tau_{max} = \max_\alpha (\tau_\alpha)$. Note that, as $C_k\to 0$, $\dot{v}_k$ approaches the finite limit  $\dot v_k \to \dot u_k - r_k \nabla^\top \dot i_k$ derived from the algebraic model (Eq. \eqref{eq:alg}) and $\dot{v}_k^\top C_k \dot{v}_k$ vanishes. Hence, only load voltages contribute to  $\dot{v}^\top C \dot{v}$ in Eq. \eqref{eq:newp}.

The matrix ${\cal Q}$ corresponding to the new BM potential ${\cal P}$ can be derived by differentiating equation \eqref{eq:newp} with respect to $x$ and using $\partial_x {\cal P} = -{\cal Q} \dot{x}$:
\begin{equation}
 	\begin{bmatrix}
     \partial_i {\cal P} \\
     \partial_v {\cal P}
    \end{bmatrix} = 
    \tau_{max}\begin{bmatrix}
     -R \dot i + \nabla \dot{v} + L\dot{i}/\tau_{max} \\
     -\nabla^\top \dot{i} - \partial_{vv}{\cal G}_0 \dot v - C \dot{v}/\tau_{max}
    \end{bmatrix}\,,
\end{equation}
\begin{equation}\label{eq:Qeq}
	{\cal Q} =
    \tau_{max}\begin{bmatrix}
	R - L/\tau_{max} & -\nabla\\
    \nabla^\top & \partial_{vv}{\cal G}_0 + C/\tau_{max}
\end{bmatrix}\,.
\end{equation}

Finally, in the limit $C_k \to 0$ on source nodes, the equation for $u$ is:
\begin{equation}\label{eq:newdudt}
    C_u \dot{u} = \partial_u {\cal P}\,.
\end{equation}

This concludes our derivation of the Brayton-Moser representation of the system dynamics in terms of the potentials ${\cal P}$ and ${\cal G}$ and matrix $Q$. These equations expose the fundamental structure of the system and will be used in the next section to analyze system stability.

\subsection{Equilibrium Point Stability}

Stability of the system can be certified by noticing that the combination of Eq. \eqref{eq:newdudt} and ${\cal Q}\dot{x} = -\partial_x {\cal P}$ with ${\cal Q}\succ 0$ can be interpreted as primal-dual dynamics with respect to the Lagrangian $-{\cal P}$ \cite{Feijer:2010ia}. The system attempts to maximize the Lagrangian with respect to the primal variable $x$ and minimize it with respect to the dual variable $u$. The dynamics converge to an equilibrium point whenever the Lagrangian is locally concave with respect to the primal variable (${\partial}_{xx} {\cal P} \succeq 0$), and convex with respect to the dual variable ($\partial_{uu}{\cal P} \preceq 0$). The second condition is automatically satisfied because ${\cal P}$ is affine in $u$, while the first condition is equivalent to $\partial_{vv} {\cal G} \succeq 0$ and can be satisfied by imposing certain constraints on the network and the operating point which are discussed below.

The asymptotic stability of the system can be formally established by considering the Lyapunov function $V$:
\begin{equation}\label{eq:lyap}
    V = \dot{x}^\top {\cal Q} \dot{x} + C_u \dot{u}^\top\dot u\,.
\end{equation} 
If there is an equilibrium point $(x^*, u^*) = (\,[i^{*\top},  v^{*\top}]^\top,\, u^*)$, V exhibits non-decreasing behavior in its neighborhood:
\begin{equation}
    \dot{V} = -\dot{x}^\top\left[\partial_{xx}{\cal P} + \dot{{\cal Q}}\right]\dot{x} \leq 0\,.
\end{equation}
Note that the matrix $\dot{{\cal Q}}$ vanishes at equilibrium and can be made arbitrarily small in its neighborhood, so the system is asymptotically stable whenever $\partial_{xx} {\cal P}(x^*, u^*) \succ 0$. Notably, these conditions can be also interpreted as the contraction behavior of the underlying dynamics \cite{Feijer:2010ia,lohmiller1998contraction} (see also example 3.7 of \cite{slotine2003modular}).

Going beyond asymptotic stability is extremely challenging. Explicit construction of the attraction regions or their approximations is beyond the scope of this work. However, a simple univariate form of nonlinearity $-p_k/v_k$ suggests that the approaches based on sector bounding of nonlinear terms, recently used for characterization of the attraction regions in ac power systems (see \cite{Vu:2015uv, Vu:2015vfb}) may be appropriate for this problem as well.

In summary, existence, feasibility and stability of at least one operating point is certified as long as the following three conditions are satisfied:
\begin{enumerate}
	\item There exist equilibrium voltage vectors $v^*, u^*$ that satisfy $\partial_v{\cal G}(v^*, u^*) = 0$ and $\partial_u {\cal G}(v^*, u^*) = 0$, and all load voltages satisfy $v_k^* > v_{\min}$. The equilibrium current vector is then given by $i^* = R^{-1}\nabla v^*$. 
    \item The Hessian of the resistive co-content is positive definite at the equilibrium point: ${\cal H}^* = \partial_{vv}{\cal G}(v^*, u^*) \succ 0$.
    \item The matrix ${\cal Q}$ is positive semidefinite at the operating point. Our definition of $\tau_{max}$ guarantees that $L -\tau_{max} R \succeq 0$, so positive semidefiniteness of $\mathcal{Q}$ reduces to $ C + \tau_{max} \partial_{vv} {\cal G}_0(v^*,u^*) \succeq 0$.
\end{enumerate}
Whenever the three conditions above hold, the stability of an operating point follows directly from LaSalle's invariance principle.

\section{Design Constraints}\label{sec:existence} 

The criteria derived in Section \ref{sec:dynamics} can be applied directly to any power system with a specified network structure. However, for ad hoc networks, these criteria must be reformulated in terms of individual components---removing the dependence on $\nabla$---which we do in this section.

We begin with a few basic assumptions:
\begin{enumerate}
\item The network graph is strongly connected. \label{cond:conn}
\item There is at least one source. \label{cond:gen}
\item The total power consumption of all the loads is bounded from above: $\sum_{k\in {\cal V}_l} p_k \leq p_\Sigma$.
\item The total resistance of all lines is bounded from above: $\sum_{\alpha \in {\cal E}} R_\alpha \leq R_\Sigma$.
\end{enumerate}

\subsection{Existence of the Equilibrium}

The equilibrium conditions for the system (Eq. \eqref{eq:main}) can be rewritten in a more traditional load-flow form:
\begin{subequations}\label{eq:equi}
\begin{align}
    p_k  = &  -v_k^*(\nabla^\top R^{-1}\nabla v^*)_k \quad k \in {\cal V}_l\,, \\
    v_k^* =& v_{\mathrm{ref}} \quad k \in {\cal V}_s\,, \\
    i_\alpha^* = & R_\alpha^{-1}(\nabla v^*)_\alpha\,.
\end{align}
\end{subequations}

Sufficient conditions for the existence of a solution to these load flow equations have been recently proposed in \cite{Bolognani:2015ek} and later extended to a number of other settings in a series of follow-up works including \cite{SimpsonPorco:2015hp,Yu:2015kl,Wang:2016vc}. It has been shown in \cite{Bolognani:2015ek} that the solution of Eq. \eqref{eq:equi} is guaranteed to exist whenever the total load is bounded by  $p_\Sigma \leq v_{\mathrm{ref}}^2/(4 \|Z\|_\infty^*)$, where $Z$ is the effective impedance matrix, which in our notation could be written as $Z = (\nabla_l^\top R^{-1}\nabla_l)^{-1}$ with $\nabla_l$ being the submatrix of $\nabla$ corresponding to load buses. The expression $\|Z\|_\infty^*$ also admits a simple interpretation. Whenever a current $(\delta j)_k$ is injected at load bus $k$, the voltage on this bus rises by $(\delta v)_k$. Then $\|Z\|_\infty^* = \max_{k\in {\cal V}_l} (\delta v)_k/(\delta j)_k$---the maximal effective resistance between the load and source buses. From the assumptions stated above, the maximum resistance between two nodes cannot exceed  $R_\Sigma$, and hence $\|Z\|_\infty^* \leq R_\Sigma$ .

This upper bound is reached when a single generator and a single load are separated by a power line of resistance $R_\Sigma$, so that is the ``worst case'' topology with respect to that condition. We conclude that, in the presence of uncertainty regarding the network structure, the solution to Eq. \eqref{eq:equi} is guaranteed to exist if and only if the maximum network load satisfies:

\begin{equation}\label{eq:exist}
    p_\Sigma \leq \frac{v_{\mathrm{ref}}^2}{4 R_\Sigma}\,.
\end{equation}

\subsection{Feasibility of the Equilibrium}
In addition to the existence of an equilibrium point, power systems typically also have a minimum voltage level requirement such that $v_k \geq v_{\min}$ on every load bus. In practical systems, this requirement (``feasibility'' of an equilibrium point) is stricter than the condition for existence of an equilibrium point. The corresponding design constraint can be derived using the expression for the voltage level:

\begin{align}
    v_k =& v_{\mathrm{ref}} - \sum_{l\in {\cal V}_l} Z_{kl}\frac{p_l}{v_l} \nonumber\,, \\
    \geq & v_{\mathrm{ref}} -  \frac{1}{\min_{l \in {\cal V}_l} v_l}\sum_{l\in {\cal V}_s} |Z_{kl}| p_l \nonumber\,, \\
    \geq & v_{\mathrm{ref}} -  \frac{1}{\min_{l \in {\cal V}_l} v_l} \|Z\|_\infty^* p_\Sigma\,.
\end{align}
Hence, we obtain a quadratic inequality for $\min_{k\in{\cal V}_l} v_k$. As long as the upper voltage solution to the load flow equations exists, it will be feasible when the following condition is satisfied:
\begin{equation}\label{eq:result2}
	p_\Sigma \leq \frac{v_{\min}(v_{\mathrm{ref}} - v_{\min})}{R_\Sigma}.
\end{equation}
This condition is more restrictive than \eqref{eq:exist} as long as $v_{\min{}} > v_{\mathrm{ref}}/2$. Inequality \eqref{eq:result2} becomes binding for the same ``worst case'' scenario as described above.

\subsection{Convexity of the BM Potential}

Next, we consider the condition for positive-definiteness of the Hessian of the BM potential at the equilibrium point: ${\cal H}^* = \partial_{vv} {\cal P}(v^*)$. 

We introduce a path-decomposition of the graph characterized by the matrix $\sigma \in \mathbb{R}^{n\times l}$ with elements $\sigma_{k\alpha} \in \{0,1,-1\}$. The values of $\sigma_{k\alpha}$ are chosen to define the path from some source bus $\kappa$ to bus $k$. Then, the voltage on bus $k$ can be represented as:

\begin{equation}\label{eq:path1}
 v_k = v_\kappa + \sum_{\alpha=1}^m \sigma_{k\alpha} (\nabla v)_\alpha\,.
\end{equation}

For convenience, we define another node in the original graph and assign it index $k = 0$. This ``virtual node'' is placed between the voltage source ($u_\kappa$) and droop resistance ($r_\kappa$) and creates an extra edge with index $\alpha = 0$ with edge impedance $R_0 = r_\kappa$. This edge connects the virtual node $s_0 = 0$ to the pre-existing node $t_0 = \kappa$. Assuming that the corresponding incidence and path-decomposition matrices are denoted as $\hat{\nabla}$ and $\hat{\sigma}$ we can write $v_\kappa = \hat{\sigma}_{\kappa 0}(\hat\nabla v)_\kappa$ and \eqref{eq:path1} can be rewritten as $v = \hat\sigma \hat\nabla v$, or

\begin{equation}
	v_k = (\hat \sigma\hat\nabla v)_k =  \sum_{\alpha=0}^m \sigma_{k\alpha} (\nabla v)_\alpha\,.
\end{equation}

The quadratic form corresponding to ${\cal H}^*$ acting on an arbitrary vector $v \in \mathbb{R}^n$ can be written:
\begin{align}
 v^\top {\cal H}^* v =&  \sum_{\alpha=1}^m R_{\alpha}^{-1}(\nabla v)_\alpha^2 + \sum_{k\in {\cal V}_s}\frac{v_k^2}{r_k} - \sum_{k\in{\cal V}_l} \frac{p_k v_k^2}{(v_k^*)^2} \nonumber\\
 =& \sum_{\alpha=0}^m R_{\alpha}i_\alpha^2 + \sum_{\substack{ k\in {\cal V}_s \\ k\neq \kappa}}\frac{v_k^2}{r_k}
- \sum_{k \in{\cal V}_l} g_k v_k^2\label{eq:Hess2}
\end{align}

We have introduced a current $i_\alpha = R_\alpha^{-1}(\hat\nabla v)_\alpha$ and an effective load conductance $g_k = p_k/(v_k^*)^2 $ that satisfies $g_k \leq p_k/ v_{\min}^2$, so the total load conductance satisfies $\sum g_k \leq p_\Sigma / v_{\min}^2$. These observations allow us to bound the last term of \eqref{eq:Hess2} with the help of the Jensen's inequality:
\begin{align}
 \sum_{k \in{\cal V}_l} g_k v_k^2 =& \sum_{k \in{\cal V}_l} g_k \left(\sum_{\alpha=0}^m \hat \sigma_{k\alpha} R_\alpha i_\alpha\right)^2 \nonumber\\
    \leq & \left(\sum_{\alpha=0}^m R_\alpha\right)\sum_{\alpha=0}^m R_\alpha i_\alpha^2 \left(\sum_{k\in{\cal V}_l} \sigma_{k\alpha}^2 g_k\right) \nonumber\\
    \leq &(R_\Sigma + r_\kappa )  \frac{p_\Sigma}{v_{\min}^2} \sum_{\alpha=0}^m R_\alpha i_\alpha^2. 
\end{align}

Therefore, the quadratic form in \eqref{eq:Hess2} is non-negative whenever the total load power and total network resistance satisfy a constraint on the maximum power demand:
\begin{equation}\label{eq:result3}
	p_\Sigma \leq \frac{v_{\min}^2}{R_\Sigma + r_\kappa}\,.
\end{equation}

Again, the ``worst case'' network configuration for this condition is the maximally separated single source and load. This condition is typically less restrictive than \eqref{eq:result2}, however it can be binding in microgrids that rely on large values of $r_k$ (larger $r_k$ values are often used for coordinated microgrid control---see Section \ref{sec:control}).

\subsection{Asymptotic Stability of the Equilibrium}

The condition ${\cal Q}(x^*) \succ 0$, with ${\cal Q}$ defined in Eq. \eqref{eq:Qeq}, can be rewritten as:
\begin{align}
\tau_{max} / r_k &> 0\,, \quad \forall k \in {\cal V}_s\,, \\
 C_k - \frac{\tau p_k}{(v_k^*)^2} &> 0\,, \quad \forall k \in {\cal V}_l\,.
\end{align}
This is always satisfied for sources. It is satisfied for loads as long as the equilibrium voltage is bounded from below by $v_k^* > v_{\min}$ and the total power consumption of the load is bounded from above:
\begin{equation}\label{eq:result1}
    p_k < C_k \frac{v_{\min}^2}{\tau_{max}}\,, \quad \forall k \in {\cal V}_l\,.
\end{equation}
As a result, there is a trade-off between load power draw and required input capacitance. In practice, $p_k$ is bounded from above by converter device ratings, so $C_k$ can be sized according to the maximum converter capacity.

In summary, we have derived three simple criteria (\eqref{eq:result2}, \eqref{eq:result3}, and \eqref{eq:result1}) which can guarantee that an equilibrium point will exist, be feasible, and be asymptotically stable for any network structure. Together, these three constraints define the set of necessary and sufficient conditions under which the arbitrary interconnection of sources, loads, and lines is guaranteed to be stable---\textit{regardless of the network topology}.

\section{DC Microgrid Control}\label{sec:control}

\subsection{Control Objectives}

The conditions derived in Sections \ref{sec:dynamics} and \ref{sec:existence} ensure that a feasible operating point exists for all admissible network topologies and loading levels. In addition to stable operation, there are two more objectives that are critical for allowing multiple sources to cooperate when connected to the same microgrid:
\begin{enumerate}
    \item each source must be able to set and update its fraction of total supplied power (``dynamic load sharing''), and
    \item node voltages $v_k$ must be maintained near the nominal network voltage $v_{\mathrm{ref}}$.
\end{enumerate}

When multiple sources are connected without coordinated control, the steady-state fraction of total power each supplies is fixed and is determined by the line resistances. This is not ideal---being able to control the relative power output of each source is a basic prerequisite for more advanced functionality, like economically-optimal power dispatch.

We assume that a vector of optimal power sharing proportions $\lambda$ has been computed from some arbitrary set of cost functions and analyze how to realize the desired power dispatch. With $\lambda_k$ defined as the participation factor that source $k$ supplies, relative to the average, the power sharing objective is given by:

\begin{equation}\label{eq:sharing}
    P_k = \lambda_k \overline{P} \equiv \frac{\lambda_k}{n_s}\sum_{l \in{\cal V}_s} P_l,\quad \forall k \in {\cal V}_s\,.
\end{equation}

At the same time, the network voltage must be regulated. To quantify this objective, we define $\overline{v} = \sum_{k \in{\cal V}_s} v_k/n_s$ as the average of the source node voltages. Our objective is to maintain this average at the desired network voltage $v_{\mathrm{ref}}$:

\begin{equation}\label{eq:avevolt}
	\overline{v} = v_{\mathrm{ref}}\,.
\end{equation}

\noindent Note that: 1) if the loads also have communication capabilities they could also be included in $\overline{v}$, and 2) much of the microgrid control literature (including \cite{GuerreroHierarchy}) considers only two-source, one-load networks when formulating their control strategies and states their objective in terms of the single load voltage: $v_l = v_{\mathrm{ref}}$. Where applicable, we have replaced $v_l$ with the more general $\overline{v}$.

\subsection{Hierarchical Microgrid Control}

To manage complexity, microgrid control is typically separated into a hierarchy of strategies operating on different timescales and with different objectives. The standard strategy \cite{GuerreroHierarchy}, which is well-known and widely-cited, is summarized below, and consists of:
\begin{enumerate}
    \item Individual (uncoordinated) control of each power converter, which, for voltage-source converters, realizes the output voltage $v_k$.
    \item Droop control, which modifies $v_k$ proportionally to the output current, mimicking a resistor. The goal of droop control is achieving power sharing.
    \item Secondary voltage control, which increases $u_k$ to offset the voltage drop caused by droop control. In its standard form, secondary voltage control only regulates the network voltage. In our proposed control strategy, it also improves power sharing.
    \item Tertiary control, which interfaces the microgrid with another power system. Tertiary control is outside the scope of this work.
\end{enumerate}

\subsubsection{Primary Control---Droop}\label{sec:droop}

When sources and loads are connected without coordinated control, the amount of power each source supplies is determined by the network configuration and line resistances. To modify the power sharing of the sources, a ``virtual resistance'' $r_k$ is inserted at the output of each converter. $r_k$ is a control parameter internal to the power electronic device and hence \textit{does not} dissipate power. For optimal performance, $r_k$ must be much greater than the line resistances so that the impact of the line resistance on the power sharing is negligible. In practice, $r_k \gg R_{\alpha}$ is typically not obtainable because larger $r_k$ values increase the node voltage deviation: $v_k = u_k - r_k (\nabla^\top i)_k$ (see Fig. \ref{fig:source} as $C_k \to 0$). This means that droop control alone cannot achieve arbitrary power sharing proportions, but it remains a popular control strategy because it is simple and does not require communication if all $r_k$ values are fixed (however, to update the power sharing proportions, the $r_k$ values must also be updated).

\subsubsection{Secondary Control---Standard Method}

To mitigate the voltage deviations caused by droop control, proportional-integral control can be used to restore the network voltage level. The standard way of achieving this is by increasing $u_k$ until equation \eqref{eq:avevolt} is satisfied:

\begin{equation}\label{eq:std_voltage}
u_k(t) = v_{\mathrm{ref}} + k_p(v_{\mathrm{ref}} - \overline{v}) + k_i \int_0^t (v_{\mathrm{ref}} - \overline{v})dt \,.
\end{equation}

In this way, all the $u_k$ values change in response to a global error signal, and under the usual assumption that $u_k(0) = v_{\mathrm{ref}}$ for all sources, all $u_k$ values are identical and the secondary voltage control does not affect power sharing.

\subsubsection{Secondary Control---Multipurpose Method}

For reasons mentioned in Sec. \ref{sec:droop}, droop control alone is not sufficient to achieve the power sharing objective (Eq. \eqref{eq:sharing}). Here, we propose a new method of secondary voltage control that ensures accurate power sharing in addition to network voltage regulation. We accomplish this by varying the $u_k$ values independently, rather than enforcing equal $u_k$ values across all sources, and combining the two control objectives into a single integral controller:

\begin{equation}\label{eq:control}
    \dot{u}_k = k_v (v_{\mathrm{ref}} - \overline{v}) + k_\lambda (\lambda_k \overline{P} - P_{k})\,,
\end{equation}
with $k_{v}, k_\lambda > 0$.

This is a simple controller that adjusts $u_k$ in response to the observed power and voltage levels. It is based on the observation that increasing $u_k$ increases both the node voltage $v_k$ and the power output of the $k$th source.

The conditions for stability of this method are still unknown; the scheme does not possess the primal-dual structure that was exploited to directly certify the stability of the uncoordinated sources described by \eqref{eq:dudt}. However, the contraction interpretation of the Lyapunov function \eqref{eq:lyap} is potentially generalizable to more sophisticated controls like \eqref{eq:control} with proper adjustment of the contraction metric. In particular, the extension of the original contraction analysis to singular systems, developed in  \cite{del2013contraction,bousquet2015contraction}, provides a natural framework for estimating the stability constraints on the controller gains $k_v$ and $k_\lambda$.

\subsection{Secondary Control: Communication Requirements}

The required communication speed and volume is an important metric when comparing microgrid control strategies. Both secondary voltage methods require all sources to share their voltage values to compute $\overline{v}$. In addition, the proposed strategy requires communication of the source currents. Accordingly, the proposed strategy requires messages which are approximately twice as long (voltage \textit{and} current of all sources) but no additional bandwidth. Further nonidealities, including communication bandwidth and delays, are not analyzed here.

\section{Microgrid Control Simulation}\label{sec:sim}

\begin{figure*}[t!]
\begin{subfigure}{.43\textwidth}
  \centering
  \includegraphics[width=\linewidth]{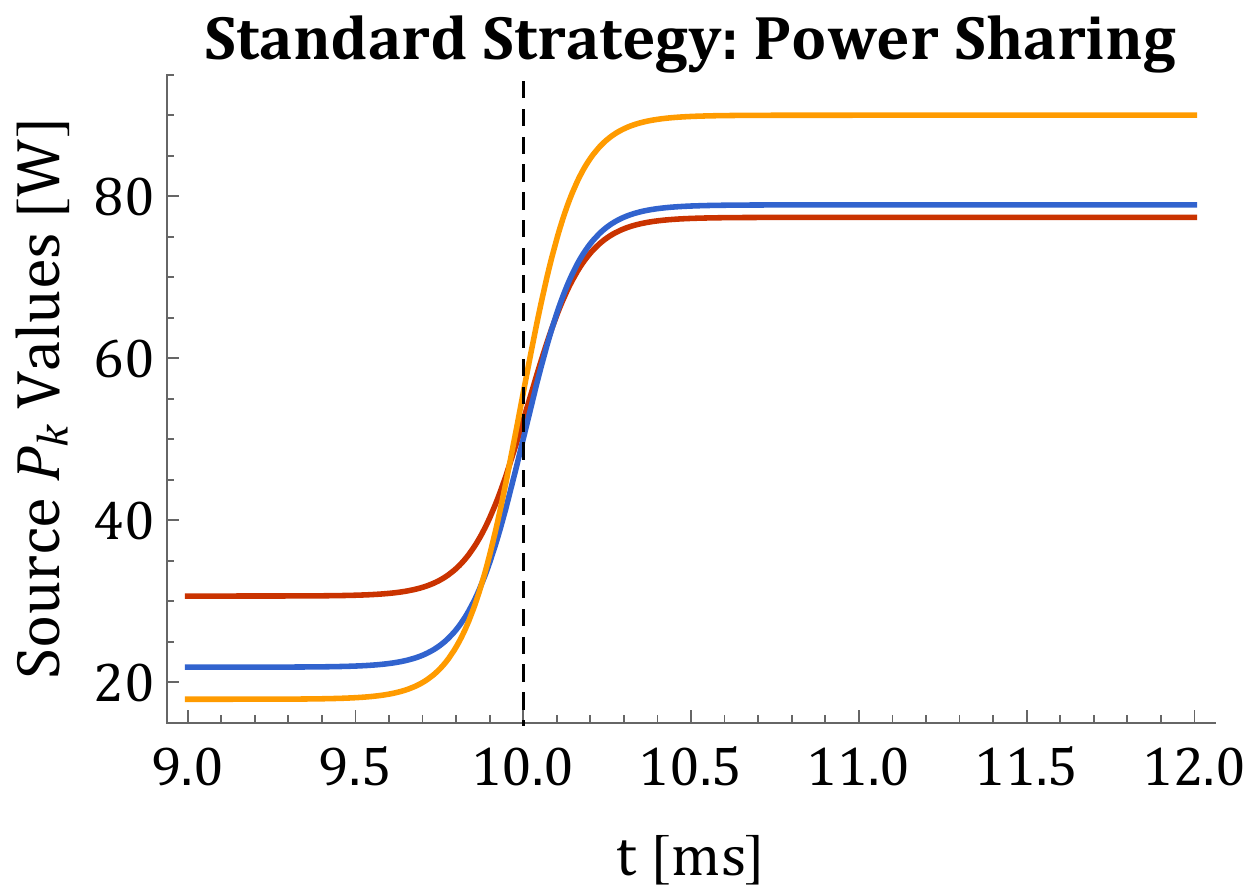}
  \caption{Droop control and standard secondary control \cite{GuerreroHierarchy}.}
  \label{fig:power_std}
\end{subfigure}
\hfill
\begin{subfigure}{.5\textwidth}
  \centering
  \includegraphics[width=\linewidth,trim={1.5cm 0 0 0},clip]{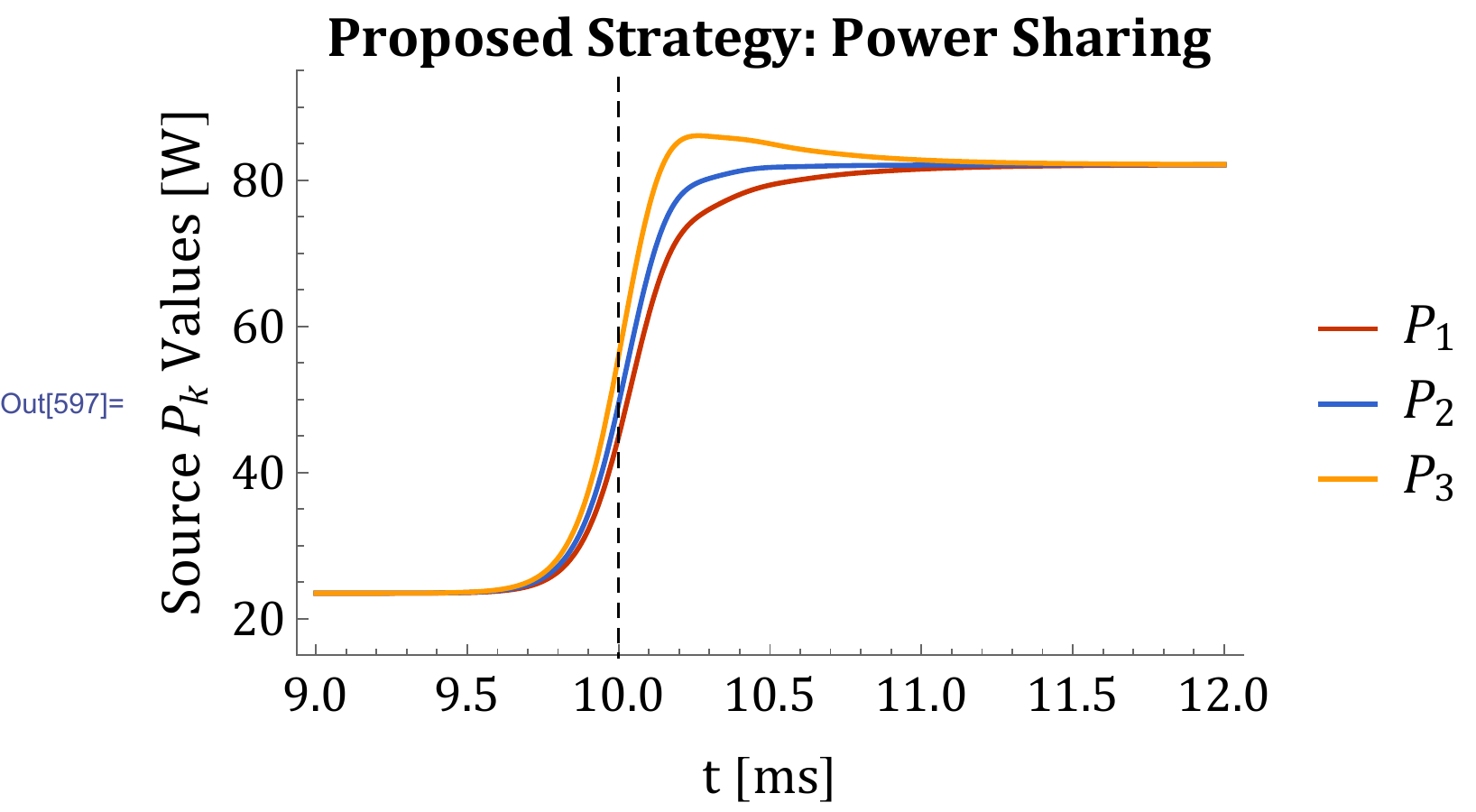}
  \caption{Droop control and multipurpose secondary control.}
  \label{fig:power_us}
\end{subfigure}
\caption{Our proposed power sharing strategy compared to the standard method of hierarchical microgrid control. Parameters chosen to achieve equal power sharing. The proposed strategy eliminates steady-state error by allowing each $u_k$ to vary independently.}
\label{fig:power}
\end{figure*}

\begin{figure}[t!]
\centering
    \includegraphics[width=0.4\textwidth]{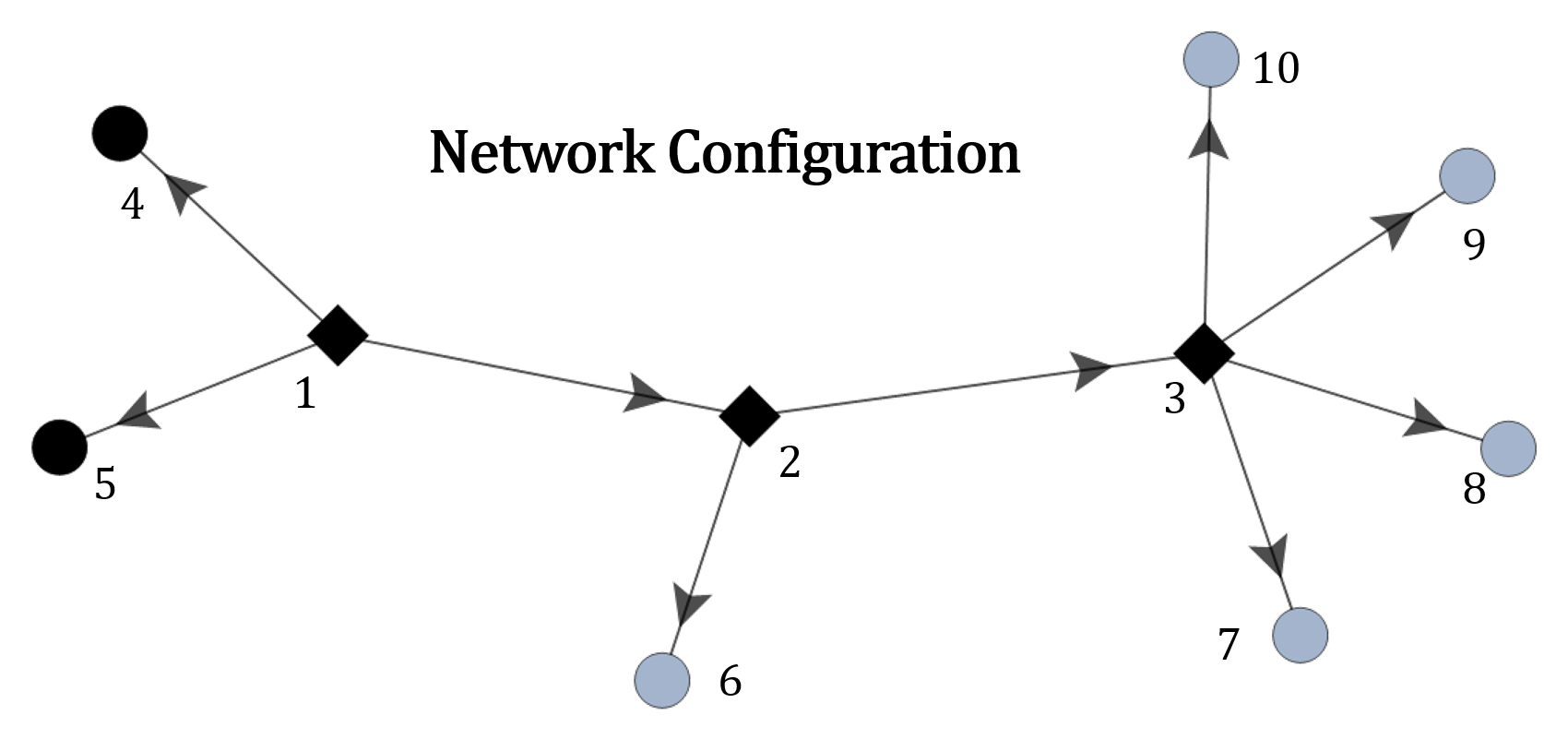}
    \caption{Visualization of the simulated network. Sources are depicted as squares, and loads as circles. Grey units turned on at \SI{10}{ms}---all other units were on throughout.}
    \label{graph}
\end{figure}

To compare our proposed secondary voltage strategy to the standard strategy, in this section we present simulation results obtained using the network configuration shown in Fig. \ref{graph} and the parameters given in Table \ref{T:Sim}. We have chosen $k_p = 0$ to simplify the comparison because we are interested in the steady-state performance. We have also done preliminary experimental validation of the method in \cite{compel}.

\begin{table}[ht!]
\caption{Simulation Parameters}
\label{T:Sim}
\begin{center}
\begin{tabular}{c | c | c}

\textbf{Parameter} & \textbf{Description} & \textbf{Value}\\ \hline

$v_{\mathrm{ref}}$ & Nominal Network Voltage & \SI{48}{\volt}\\
$R_{\alpha}$ & Line Resistance & \SI{0.111}{\ohm} \\
$\tau_{\alpha}$ & Line Time Constant & \SI{55.45}{\micro\second}\\
$r_{k}$ & Droop Resistance & \SI{0.5}{\ohm} \\
$C$ & Load Input Capacitance & \SI{845.7}{\pico\henry}\\
$p_k$ & Load Power & \SI{35.11}{\watt}\\
$k_p$ & Proportional Gain & \SI{0}{} \\
$k_i$ & Integral Gain & \SI{18.02}{\per\second}\\
$k_{v}$ & Voltage Error Gain & \SI{36.04}{\per\second} \\
$k_{\lambda}$ & Power Sharing Error Gain & \SI{0.7508}{\per\ampere\per\second} \\
$\lambda_k$ & Participation Factor of Source $k$ & $1$ \\

\end{tabular}
\end{center}
\end{table}

\begin{figure}[h!]
\centering
        \includegraphics[width=0.42\textwidth]{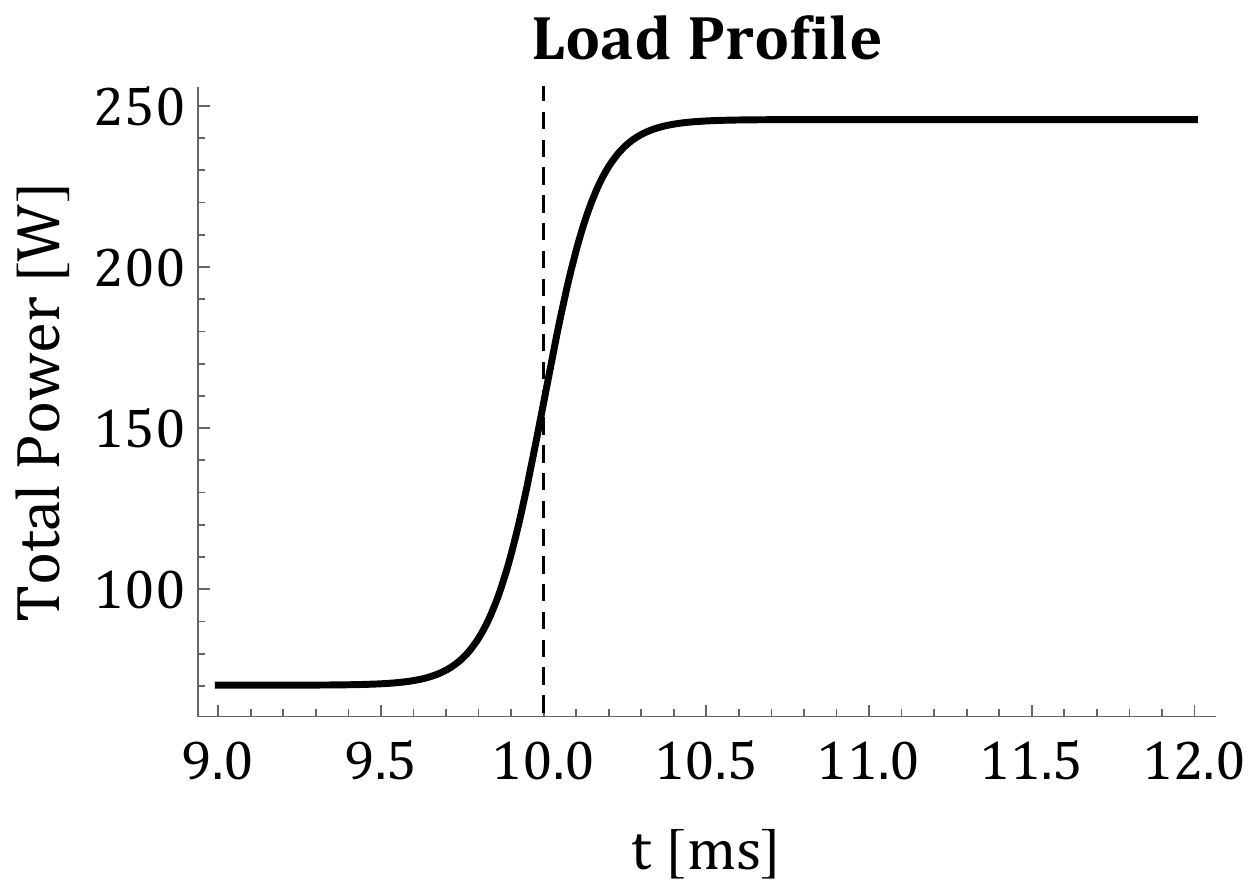}
        \caption{Total power drawn by all loads in the network.}
        \label{loadprofile}
\end{figure}

\begin{figure}[ht!]
  \centering  
  \includegraphics[width=0.49\textwidth,trim={1.7cm 0 0 0},clip] {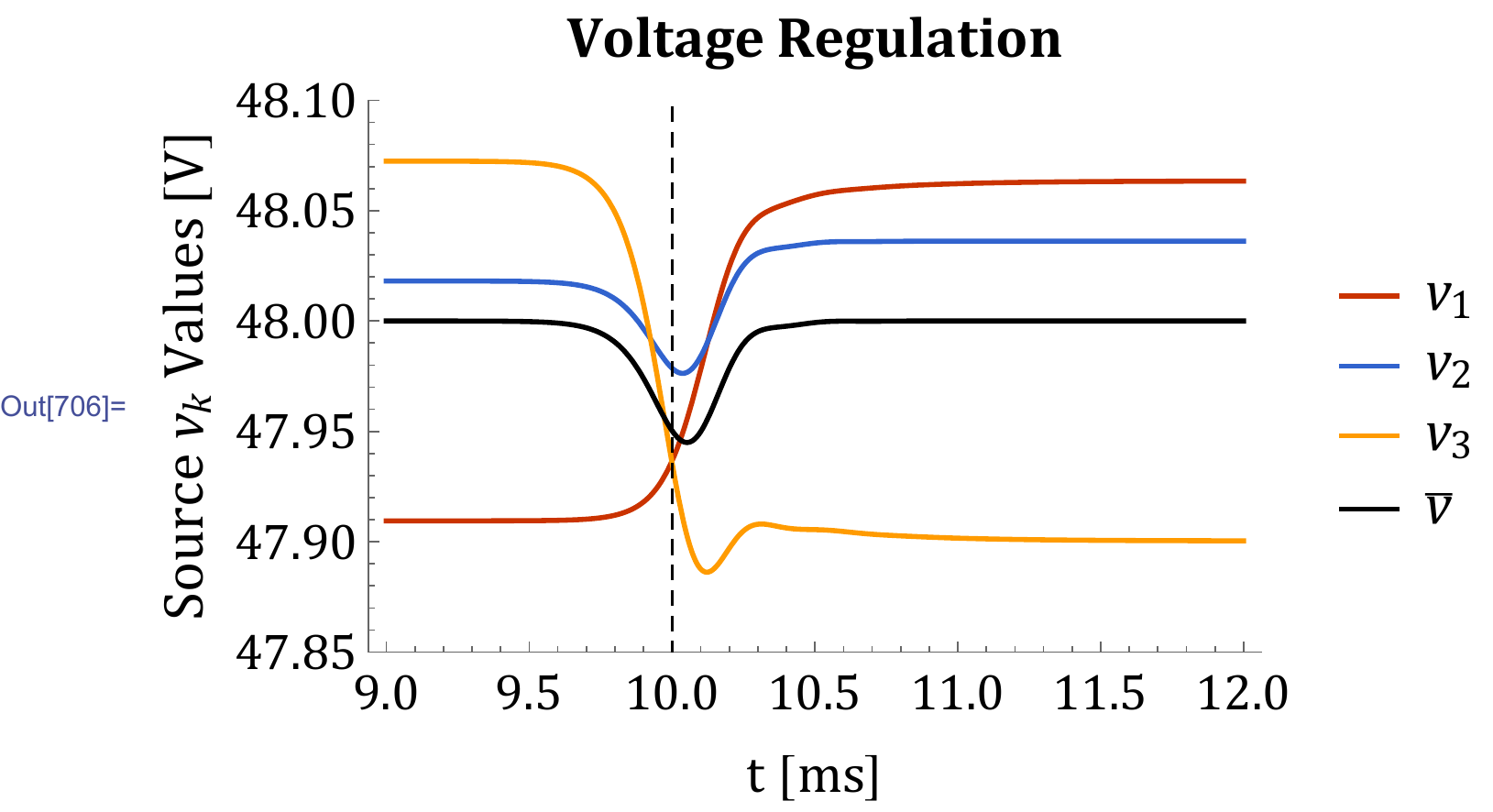}
  \caption{Voltage regulation capability of the proposed strategy.}
  \label{fig:vreg}
\end{figure}

The network has a total of seven loads---initially, two were on, and the other five turned on at \SI{10}{ms} as shown in Fig. \ref{loadprofile}. Source, load, and line models are as depicted in Fig. \ref{fig:models}. The power sharing results obtained by the standard voltage restoration strategy and our strategy are shown in Fig. \ref{fig:power}. The droop resistances and $\lambda_k$ values were designed for all sources to share equally ($P_1 = P_2 = P_3$). The proposed strategy obtains precise power sharing because it uses the secondary voltage control for power sharing in addition to voltage regulation. The standard strategy has an unavoidable steady-state power sharing error because it uses only proportional primary (droop) control. Both approaches achieve comparable voltage regulation (results for the standard strategy are not shown), and the source node voltages in the proposed strategy are shown in Fig. \ref{fig:vreg}.

\section{Conclusions and Path Forward}

Ad hoc microgrids present an alternative to the inherently centralized nature of bulk power systems. By allowing nonspecialist consumers to safely configure and maintain their own electricity infrastructure and markets, they have unparalleled potential to provide power to communities underserved by traditional power systems. In the past, development of ad hoc networks has been impeded by the difficulty of guaranteeing the stability of arbitrary interconnections of power electronic subsystems a priori. In this paper we have relied on a number of classical control theory techniques, including Brayton-Moser potentials, primal-dual dynamics, and recent approaches to the analysis of load flow equations, to develop conditions under which ad hoc dc microgrids are stable. In particular, we have developed conditions on the overall system power consumption (Eqs. \eqref{eq:result2} and \eqref{eq:result3}) and individual load capacitance values (Eq. \eqref{eq:result1}) such that the stability of arbitrary interconnections of voltage-source converters and tightly-regulated power electronic loads is guaranteed.

In addition, we have proposed a new decentralized control strategy for autonomous coordination of many sources to achieve dynamic load sharing while regulating the network voltage. As shown in our simulated comparison, this strategy obtains more accurate power sharing than the decentralized microgrid control techniques used in practice. 

Finally, there are several exciting open questions that still need to be answered:
\begin{itemize}
    \item What are admissible and optimal values of the decentralized control gains that ensure system stability and performance in ad hoc microgrids?
    \item Can more flexible stability conditions be derived by adding carefully chosen constraints to the network topology---for example, by focusing on a particular class of networks?
    \item What are the effects of practical limitations (like communication delays and outages) on decentralized controller performance?
    \item How can the transient stability of ad hoc systems be characterized?
    \item How resilient are ad hoc systems to disturbances caused by routine switching events and/or emergency faults?
\end{itemize}

In this paper, we have addressed some challenges inherent to ad hoc microgrids and hope that these challenges continue to be resolved, with the ultimate goal of enabling the large-scale deployment of truly decentralized power systems.

\bibliographystyle{ieeetr}

\end{document}